\documentclass[%
 reprint,
 amsmath,amssymb,
 aps,
floatfix,
]{revtex4-2}

\usepackage{graphicx}
\usepackage{dcolumn}
\usepackage{bm}
\usepackage{comment}

\newcommand{\expect}[1]{\left\langle {#1} \right\rangle}

\begin{document}

\preprint{APS/123-QED}

\title{Proposal for Majorana Modes without a Magnetic Field in a semiconductor-superconductor sandwich structures}

\author{Huan-Kuang Wu}
\author{Jay D. Sau}%
 \email{jaydsau@umd.edu}
\affiliation{%
 Department of Physics, Condensed Matter Theory Center and Joint Quantum Institute,
University of Maryland, College Park, Maryland 20742, USA}%
\date{\today}
\begin{abstract}
We propose a planar Josephson junction setup to host Majorana modes where the semiconductor is sandwiched between superconductors on both 
surfaces.
While the studied configuration is related to both devices proposed for use with topological insulators as well as other proposals for Majorana modes without    applied magnetic fields, we find that placing the superconductor on both surfaces allows us to replace the topological insulator by a wider variety of spin-orbit coupled semiconductors.
Application of an electric field in the junction together with appropriate superconducting phase differences allows different Fermi surfaces to be 
subject to different phase differences. 
We find that these conditions can drive the system into a class-D topological superconductor with a pair of Majorana modes at the two ends of the junction.
Additionally, class-DIII topological superconductors with helical Majorana modes also occur in other parts of the phase diagram.
We simulate our setup on HgTe near the valence band edge described by the well-known 8-band Kane model with realistic parameters and find 
a large topological regime in the 
phase diagram can be achieved.  The topological gap can remain comparable to the bulk superconducting gap throughout the fermi surface.
\end{abstract}

\maketitle


\section{\label{sec:1}Introduction}

Topological superconductors (TSC) have attracted considerable interest in condensed matter physics owing to their potential application to topological quantum computation~\cite{kitaev2001unpaired, kitaev2003fault, freedman2003topological, sarma2005topologically, nayak2008non}. 
This phase of matter was first predicted in  spinless p-wave superconductors~\cite{read2000paired,kitaev2001unpaired,sengupta2001midgap} and 
is characterized by the occurrence of localized zero-energy Majorana zero modes (MZMs). 
Since then, a large number of designs have been proposed and studied~\cite{qi2011topological, sau2010generic, sau2010non, lutchyn2010majorana, oreg2010helical,alicea2012new, leijnse2012introduction, bernevig2013topological, beenakker2013search} for the realization of MZMs in real physical systems. 
Many of these proposals~\cite{sau2010generic,lutchyn2010majorana,oreg2010helical} support the TSC phase in a limited range of chemical 
potential, which can be difficult to control in systems that are in contact with a superconductor.
This problem is avoided for the TSC phase at the interface of a strong topological insulator in proximity to an s-wave 
superconductor~\cite{fu2008superconducting} where a vortex is known to support an MZM.
However, the MZMs in most of these systems is accompanied by a large number of closely spaced Andreev states, 
making it difficult to unequivocally identify the MZM~\cite{bai2020novel, bai2022proximity,endres2022transparent,flototto2018superconducting}.
Another more recent class of proposals where the chemical potential constraints are relaxed is the so-called planar Josephson junction~\cite{pientka2017topological, hell2017two}, which is composed of a Rashba spin-orbit coupled 2DEG between two s-wave superconductors with a phase difference $\phi$. 
Such a junction is gapless at phase $\phi=\pi$ and the application of an in-plane magnetic field opens up a topological regime in the vicinity of $\phi\sim\pi$.
Recent experiments~\cite{fornieri2019evidence, ren2019topological, banerjee2022signatures, banerjee2022local} on the planar Josephson junction have observed signatures of the topological phase transition.

Most of the proposals described above generate time-reversal asymmetry required for isolated MZMs using a magnetic field.
Such magnetic fields in combination with disorder often leads to degradation of the topological gap, which is determined by the superconductivity~\cite{tinkham2004introduction}.
Therefore, the appearance of MZMs in this system requires careful control of the disorder~\cite{takei2013soft}. One exception to this is the MZMs generated in 3D topological insulator(TI)/superconductor interfaces where the MZM can occur in a 
vortex confined to a tri-junction~\cite{fu2008superconducting}. Since then several 
proposals for obtaining isolated Majorana zero modes using a phase bias have been suggested both in nanowires~\cite{cook2011majorana,romito2012manipulating,de2019conditions,vaitiekenas2020flux} and planar Josephson junction setups~\cite{lesser2022majorana}.
Other designs include biasing the phase of the superconductors using a supercurrent driven  parallel to the Josephson junction~\cite{melo2019supercurrent} as well as an SNSNS junction~\cite{lesser2022one} with the three superconductors at specific phase biases. However,
the SNSNS requires using a semiconductor with a spin-dependent Fermi velocity~\cite{lesser2022one}.
An interesting feature of all these proposals is that they effectively require 
multi-terminal Josephson junction setups with effectively three superconductors~\cite{van2014single}.



\begin{figure}
    \centering
    \includegraphics[width=0.48\textwidth]{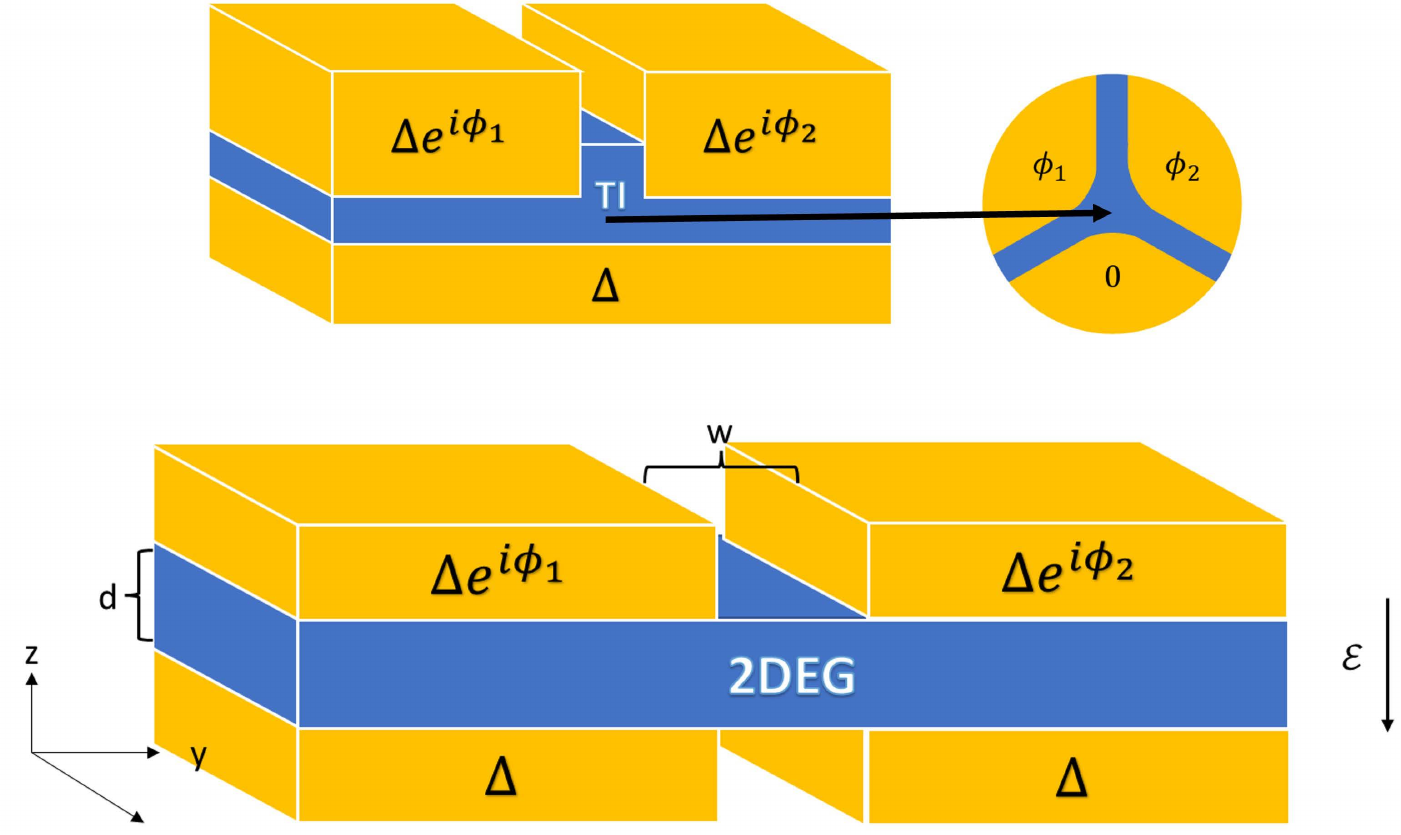}
    \caption{(Top panel) Schematic for topological superconducting JJ in a 3d TI~\cite{fu2008superconducting}, where a tri-junction is formed on the side surface by superconductors on the top and bottom. An MBS can appear at the intersection of the tri-junction when the three phases $(0, \phi_1, \phi_2)$ form a vortex. (Bottom panel) The device can remain topologically non-trivial even if the 3d TI is replaced by a spin-orbit coupled 2DEG. In this device, four s-wave superconductors are in contact with a quasi-2d spin-orbit coupled semiconductor. An external electric field $E$ points to the $-z$ direction. 
    }
    \label{fig:1}
\end{figure}

Recent advances in flip-chip fabrication techniques 
have made it possible to create devices with superconductors below a semiconductor layer~\cite{flototto2018superconducting}, in addition to superconducting layers that are standard to deposit on top of the semiconductor. Motivated by these works
, we propose a family of devices (see Fig.~\ref{fig:1}) for MZM without a Zeeman field that takes advantage of placing superconductors above 
and below the superconductor. Additionally, this structure provides a natural interpolation between a 3D TI thin film to a
spin-orbit coupled semiconductor. We will find that the device also support a class DIII topological superconducting phase~\cite{zhang2013time}
at some of the boundary of the phase with MZMs.
Our system is controlled by three s-wave superconductor phases $(0, \phi_1, \phi_2)$~\cite{van2014single}, the first one below the semiconductor and 
the other two form a Josephson junction on the top surface. 
An external electric field applied in the vertical direction, is used to control the strength of the proximity effect at the top and bottom interface.
Utilizing the inequality in the Cooper pair tunneling amplitude across the boundary for the two Fermi surfaces, the system can be tuned into a class D topological superconducting phase~\cite{altland1997nonstandard,schnyder2008classification}, where a pair of Majorana zero modes appear at the two ends of the normal region in the middle. 
When the phase bias $\phi_{1,2}$ at  the top surface (relative to the bottom) is at a time-reversal symmetric point ($0$ or $\pi$), the 2DEGs on either side of the JJ becomes  two-dimensional class DIII systems, whose $\mathbb{Z}_2$ topological invariant~\cite{schnyder2008classification, kitaev2009periodic, qi2010topological} corresponds to the parity of the number of Fermi surfaces with a negative pairing potential that encloses a time-reversal invariant momentum~\cite{ teo2010topological, zhang2013time}.
In the parameter regime where the proximity effect is stronger on the opposite interfaces for the two Fermi surfaces, the system is topologically non-trivial at $\phi_{1,2} = \pi$, hosting helical Majorana edge modes at the boundary.
Note that the mechanism for class DIII topological superconductivity is similar to the one-dimensional system proposed in Ref.~\cite{keselman2013inducing}, but here an electric field is required to break the mirror symmetry in the z direction explicitly.

The manuscript is organized as follows. In Sec.~\ref{sec:2}, we describe the structure of our setup, whose model is introduced in Sec.~\ref{sec:3}. The topological phase boundary of our system is derived and presented in Sec.~\ref{subsec:3C}. In Sec.~\ref{subsec:3D}, the topological phase boundary in the case where the Cooper pair tunneling is stronger on the opposite sides for the two Fermi surfaces is discussed. In Sec.~\ref{sec:4}, we demonstrate the topological superconductivity by simulating our setup using HgTe as the 2DEG with realistic parameters and the topological gap, which is the lowest Andreev bound state energy, is shown in Sec.~\ref{subsec:4D}.

\section{Analogy with 3D TI trijunctions}
\label{sec:2}
To understand the conditions under which the set-up in Fig.~\ref{fig:1} can support MZMs, let us first review the tri-junction proposal in TIs~\cite{fu2008superconducting} where superconductors are deposited to cover all faces of a thick TI. 
The uncovered regions on the sides and the top face of the TI form line Josephson junctions 
where the Josephson current is carried by the surface state. 
These line junctions intersect to form a tri-junction on the side surface.
Such a tri-junction traps a Majorana bound state (MBS) when the three phases $\phi_1, \phi_2$ and $\phi=0$ (bottom superconductor) forms a vortex~\cite{fu2008superconducting}.

An alternative understanding for the emergence of the MZMs, which is more useful in generalizing to thin films TIs and 
2DEGs (i.e. both panels of Fig.~\ref{fig:1}), is to view the system as a planar Josephson junction~\cite{pientka2017topological}. The emergence of topological superconductivity 
in these systems relies on the fact that a high transparency JJ in an SC in the Andreev limit (i.e. $\mu\gg\Delta_{i,\pm}$) supports bound ABSs with a gap that closes when the phase difference across the JJ crosses $\pi$~\cite{tinkham2004introduction} (see appendix ~\ref{app.B} for a review). A planar 
Josephson junction, under the same conditions, would support a gap closing and re-opening at $k_x=0$, when the phase difference $(\phi_1-\phi_2)$ crosses $\pi$. Here $k_x$ is the momentum of ABSs along the planar JJ on the top surface of the device shown in Fig.~\ref{fig:1}. In contrast to conventional 2DEGs, the surface state of the TI is non-degenerate and only one pair of ABSs crosses zero at $k_x=0$. 
This crossing of ABSs in a TI JJ for a phase difference of  $(\phi_1-\phi_2)=\pi$ is consistent with the prediction of helical Majorana modes 
under these conditions~\cite{fu2008superconducting}.

These conclusions may be shown to robust to breaking the Andreev approximation or finite transparency of the junction by considering the change 
of the $Z_2$ topological invariant across such a crossing of ABSs. In fact, the zero-energy crossing of a single pair of ABSs at $k_x=0$ in a 
one dimensional system is known to necessarily change the $Z_2$ topological invariant of a system~\cite{kitaev2001unpaired,read2000paired}. Therefore, 
one side of $(\phi_1-\phi_2)=\pi$ must be a non-trivial class $D$ topological superconductor with an invariant of $-1$. Introducing a finite back-scattering or other forms of time-reversal symmetry breaking can shift the crossing point, but cannot eliminate the TSC phase. Incidentally, in the case of a thick 
3D TI slab, time-reversal symmetry is preserved when $(\phi_1-\phi_2)=\pi$, so that the introduction of backscattering at the interface cannot shift the 
the ABS zero energy crossing topological phase transition point.

The TSC phase is characterized by the appearance of MZMs at the ends, which happen to be on the side surface of the device in Fig.~\ref{fig:1}. 
Note that the 3D TI slab has a second fermi surface on the bottom surface of the TI. This surface, unlike some other proposed TI/SC devices~\cite{Potter2013,Kurter2015}, is assumed to be gapped by a bottom superconductor with no phase 
difference and therefore doesn't contribute. This is what allows the device shown in Fig.~\ref{fig:1} to support topological MZMs without a 
magnetic field.
In contrast, the conventional planar JJ proposals~\cite{pientka2017topological} for realizing MZMs require a parallel magnetic field 
to shift the closing of the ABS gap in the second Fermi surface away from $(\phi_1-\phi_2)=\pi$ to create a window where one Fermi surface is 
topologically non-trivial while the other one remains trivial. Since TSCs in symmetry class $D$ only support a $Z_2$ topological invariant, the 
phase where both fermi surfaces are topologically non-trivial is equivalent to a trivial phase when Fermi surface coupling is considered.


\section{2DEG/SC sandwich structures}
\label{sec:3}

The central question that we are interested in is whether the bulk 3D TI in Fig.~\ref{fig:1} can be replaced by a spin-orbit coupled 2DEG. 
In fact, a TI slab is technically a quasi two-dimensional structure and one expects the vortex Majorana modes to be robust to a small overlap between surface states. A spin-orbit coupled 2DEG is similar to a TI slab with a large overlap between the surface states in the sense that both two dimensional 
Fermi surfaces are delocalized over the thickness of the 2DEG.

The key features that allowed the 3D TI discussed in the previous subsection to support a TSC phase was that the Fermi surfaces were decoupled and
one of the Fermi surfaces had a stronger overlap with the JJ on the top surface compared to the other Fermi surface. The same argument can be applied to any 2DEG where, as we will show in the next section, an electric field can be used to create a similar spatial as well as momentum separation of the 
Fermi surfaces. 

\subsection{Spatial/momentum splitting of Fermi surfaces}
\label{subsec:3A}
The presence of a $S_4$ mirror symmetry in the $x-y$ plane of the 2DEG would map $z\rightarrow -z$ that would rule out a difference in weight between 
the top and bottom surfaces of the 2DEG. This mirror symmetry can be broken by  
an external electric field that is assumed to be applied in the $-z$ direction. 
The separation between the bottom superconductors  in Fig.~\ref{fig:1}
allows for the possibility of dual gating where the electric field and electron density in the junction can be controlled independently. 
In addition to the asymmetry in weight between the top and bottom surfaces(i.e. spatial separation), the electric field 
generates a Rashba spin-orbit coupling that leads to a splitting of the 
Fermi surfaces in momentum space~\cite{winkler2003spin}. 

A description of Rashba spin-splitting arising from an electric field requires analyzing a multiband Hamiltonian. 
We avoid this complication in this section, for purposes of illustration of this separation, we consider a quantumm well with Dresselhaus spin-orbit coupling 
that is described by a Hamiltonian:
\begin{align}
H = H_0 + H_D
\end{align}
with $H_0 = (p_x^2+p_y^2+p_z^2)/2m + V(z)$ where $V(z)$ is the confinement potential $V(z) = e\mathcal{E}(z-d/2)$ for $0\leq z\leq d$ and $\infty$ otherwise, where $\mathcal{E}$ is an applied electric field. $H_D$ represents a Dresselhaus spin-orbit interaction
\begin{equation}
    H_D = D[ p_x(p_y^2-p_z^2)\sigma_x + p_y(p_z^2-p_x^2)\sigma_y + p_z(p_x^2-p_y^2)\sigma_z].
    \label{eq:full_dresselhaus}
\end{equation}
Note that such a model describes 2DEGs with bulk inversion asymmetry as well as continuous rotation symmetry breaking,
which is not a necessary ingredient for the spatial and momentum separation.
To simplify the calculation, let us first consider the momentum along the direction $p_x=p_y=p$. Along this direction, the 
Hamiltonian commutes with $(\sigma_x-\sigma_y)$, so that we can assume the eigenstates to be in spin-states $(\sigma_x-\sigma_y)=\pm \sqrt{2}$.
The Hamiltonian in each of these sectors is written as 
\begin{align}
    &H_\pm = p^2/m+p_z^2/2m\pm D\sqrt{2}(p^2-p_z^2)p+V(z).
\end{align}
As seen from the appendix~\ref{app.0}, the eigenvalues to lowest order in $p$ and $D$ is written as 
\begin{equation}\label{eq:disp}
\varepsilon_\pm(\vec{p}) = \frac{p^2}{2m} \pm \frac{\beta}{\hbar} p,
\end{equation}
where the linear in $p$ Dresselhaus spin splitting~\cite{winkler2003spin} is proportional to $\beta=\sqrt{2}\alpha\mathcal{E}^{2/3} Dp/3$ is generated by the 
$z$ confinement by the electric field, which creates a non-zero $\expect{p_z^2}$ in Eq.~\ref{eq:full_dresselhaus}.
The Hamiltonian $H$ is rotation invariant up to linear order in $p = \sqrt{p_x^2+p_y^2}$, therefore this spin splitting is 
rotationally invariant and the corresponding Fermi surfaces are separated in momentum space. Note that this splitting is in addition to 
Rashba spin splitting that is also expected to arise in the presence of an external electric field.

Let us now discuss the spatial distribution of the wave-function along the $z$ direction. Note in the above calculation. for simplicity, we have ignored the finite width $d$ of the quantum well. However, the superconducting proximity effect from the top and bottom surfaces is proportional to the tunneling amplitude $|\phi'(0)|^2$ and $|\phi'(d)|^2$. Following the solution in the appendix~\ref{app.0} the relative tunneling at the bottom surface and the top surface are written as:
\begin{align}
 &\epsilon=|\phi'(d)|^2/|\phi'(0)|^2=4 Ai'(\lambda^{-1}d-\nu_2)^2/Ai'(-\nu_2)^2\label{eq:epsilon},
\end{align}
where $\lambda^3=\mathcal{E}^{-1}(1/2m\mp D\sqrt{2}p)$. The factor of $4$ in the latter equation is discussed in the appendix~\ref{app.0} and  arises from the boundary at $d$ in the limit that the confinement is dominated by the electric field. Note that the ratio of the weight is different for the different 
Fermi surfaces associated with the spin directions $\pm$ are different because of the difference in  $\lambda$ from the sign. Here the radius of the Fermi surface $p\sqrt{}2\simeq \sqrt{2 m\mu}$, where $\mu$ is the chemical potential. Physically, the difference in the weights arises from the fact that the 
$p_z^2$ term in Eq.~\ref{eq:full_dresselhaus} can be thought of as a spin-dependent contribution to the effective mass. This, in turn, changes the extent 
of confinement by the electric field. 

While the Dresselhaus model discussed in this section provides a simple model to understand the electric field induced separation of the spin-split 
Fermi surfaces in momentum $p$ as well as between the top surfaces, the effect is relatively weak in the single-band model considered here. We will see from a numerical solution discussed in the next section, that these effects are significantly enhanced in the 8-band Kane model for HgTe~\cite{winkler2003spin}.

\subsection{Effective pairing potential from the two surfaces}
\label{subsec:3B}
To understand the role of the spatial separation of the different Fermi surfaces  consider one side of the Josephson junction as in Fig.~\ref{fig:1} with a pair of s-wave superconductors attached to the top and bottom surfaces of the 2DEG. 
We will assume that the momentum splitting resulting from the dispersion in Eq.~\ref{eq:disp} is large enough so that the scattering between Fermi surfaces during Andreev reflection at the JJ in Fig.~\ref{fig:1}
may be ignored. As a result we will treat the ABS from each Fermi surface in isolation, similar to the TI analysis in the previous sub-section.
We assume phase bias $\phi$ for the superconductor at the top surface and phase $0$ for the superconductor at the bottom surface. 
The proximitized pairing potential across an SN interface is determined by the tunneling amplitude from the superconductor into the 2DEG~\cite{stanescu2010proximity}, which depends on the partial derivative of the wave function in the $z$ direction at the boundaries~\cite{chen1988theory},
which was calculated in Eq.~\ref{eq:epsilon}. 
Specifically, denoting the profile of the 2DEG electrons by $\varphi(z)$, the pairing potential contributed by each of the superconductors on the top and bottom are written as $\Delta \propto \left\vert \partial_z\varphi(z=z_0)\right\vert^2$ where $z_0\in\{0,d\}$.
In the limit $d\ll \xi$, the effective pairing potential is simply the sum of contributions from the top and bottom superconductors
\begin{equation}
    \Delta_{\text{eff}} = \Delta_b + \Delta_t
\end{equation}
where
\begin{align}
    \Delta_t &=\Delta_0 d^3\left\vert \partial_z\varphi(z=0)\right\vert^2 \nonumber\\
    \Delta_b &=\Delta_0 d^3\left\vert \partial_z\varphi(z=-d)\right\vert^2=\epsilon \Delta_t
\end{align}
and $\Delta_0$ is a nominal pairing amplitude (apart from a factor of length) of the semiconductor/superconductor interfaces.
Note that the factor of $d$ is introduced to ensure that $\Delta_0$ has dimensions of energy. 
Assuming both superconductors has the same interface transparency, the effective pairing potential can be expressed as
\begin{equation}
    \Delta_{\text{eff}} = \Delta_t\left(\epsilon + e^{i\phi}\right).
    \label{eq:effective_pairing}
\end{equation}
As evident from Eq.~\ref{eq:epsilon}, the factor $\epsilon$ and therefore the phase of the effective pairing potential $\Delta_{eff}$ 
depends on the spin index $\pm$ of the spin-split Fermi surface. Thus, the confinement induced momentum and vertical separation of 
the Fermi surface wave-functions can allow us to independently control the effective superconducting phase in each Fermi surface.

\subsection{Topological phase boundaries}
\label{subsec:3C}
The quasi 1-d Josephson junction between the two superconductors in Fig.~\ref{fig:1} exhibits particle-hole symmetry but no time-reversal symmetry and is therefore in class D. 
Therefore, similar to the JJ discussed in Sec.~\ref{sec:2}
the topological invariant changes whenever the gap in the Andreev spectrum at $k_x=0$ closes and reopens as the phases $\phi_{1,2}$ are varied.
appears in the band structure. 
Following Eq.~\ref{eq:effective_pairing}, the effective pairing potential  on each of the sides $i\in\{1,2\}$ of the JJ in Fig.~\ref{fig:1} for the $\pm$ Fermi surfaces is written as: 
\begin{equation}
    \Delta_{i,\pm} = \Delta_t\left(\epsilon_\pm+e^{i\phi_i}\right)\label{eq:Dpm}
\end{equation}
where $\epsilon$ is dependent on  the z-dependent part of the wave function at $k_x = 0$ for the two Fermi surfaces by  Eq.~\ref{eq:epsilon}. 

As mentioned in Sec.~\ref{sec:2} (see appendix ~\ref{app.B} for details), the gap of ABSs at $k_x=0$ in a JJ in the Andreev limit ($\mu\gg\Delta_{i,\pm}$) 
closes when the phase difference across the junction becomes $\pi$. 
This corresponds to the condition
\begin{align}
    \frac{\epsilon_{\pm} + e^{i\phi_1}}{\vert \epsilon_\pm + e^{i\phi_1}\vert} = -\frac{\epsilon_\pm + e^{i\phi_2}}{\vert \epsilon_\pm + e^{i\phi_2}\vert},
    \label{eq:band_inversion}
\end{align}
for each of the Fermi surfaces $\pm$. The TS phase with MZMs would exist between the two points in phase where $\epsilon_\pm$ satisfy the above equation.
Note that a necessary condition to satisfy Eq.~\ref{eq:band_inversion} is $\epsilon<1$.
\begin{figure}
    \centering
    \includegraphics[width = 0.45\textwidth]{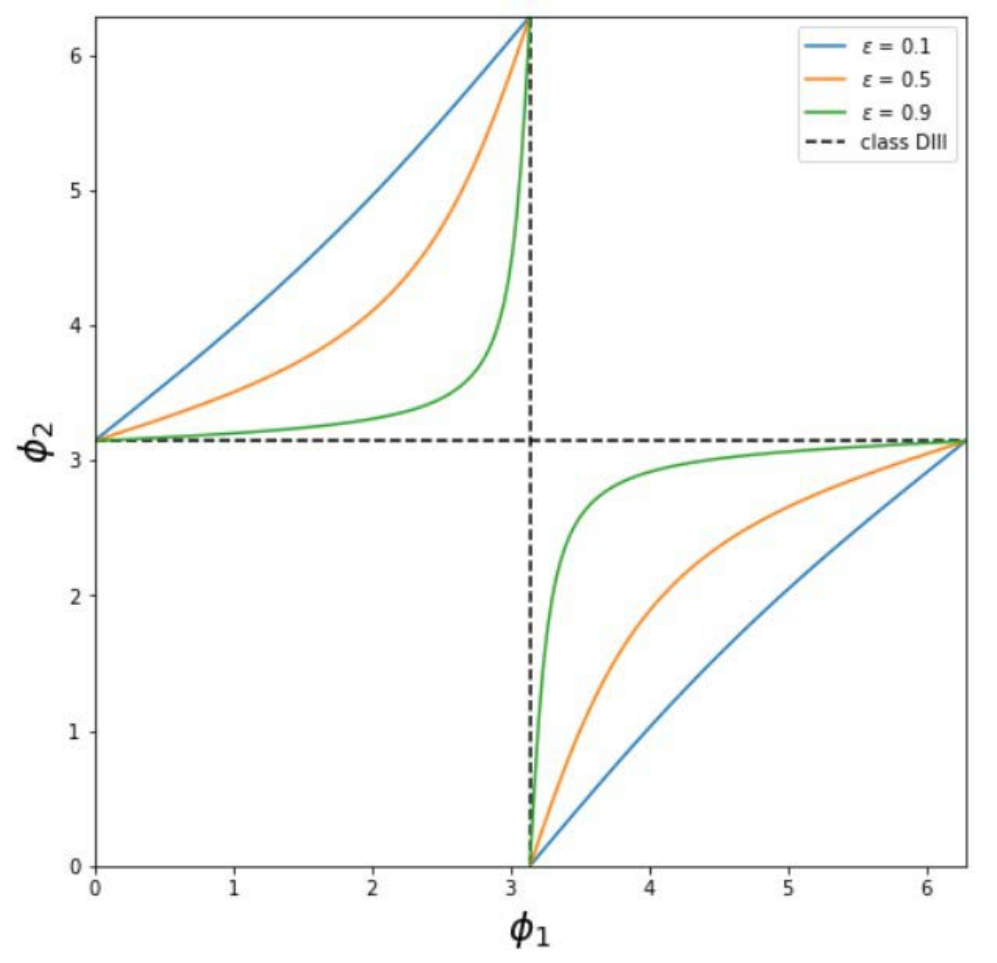}
    \caption{Gap-closing curves in the $\phi_1, \phi_2$ space for different $\epsilon$. The dashed line corresponds to the helical Majorana states in the class DIII limit at $\phi_{1,2} = \pi$ }
    \label{fig:2}
\end{figure}

To understand the phase diagram of the TS phase of the 2DEG shown in the bottom panel of Fig.~\ref{fig:1}, in the Andreev limit, we plot the gap-closing curves in Eq.~\ref{eq:band_inversion} for different $\epsilon<1$ in Fig.~\ref{fig:2}.
As described in Sec.~\ref{subsec:3A}, the combined effect of the external electric field and spin-orbit coupling can lead to different values of 
$\epsilon_\pm$ for the two Fermi surfaces with spinlabels $\pm$.
Consequently, as shown in Fig.~\ref{fig:2}, there will be two regimes in the $\phi_1, \phi_2$ space enclosed between the two curves that correspond to $\epsilon_+ < 1$ and $\epsilon_-<1$.
Across the boundaries, only one band inversion takes place 
, resulting in a topological phase transition.
Since the system is known to be topologically trivial at phases $\phi_{1,2} = 0$, the two enclosed regime between these curves is a TS phase with end MZMs.

\subsection{Gapless Majorana edge mode in the class DIII limit}
\label{subsec:3D}
The result in the previous subsection i.e. Eq.~\ref{eq:band_inversion} makes it clear that there is no TS phase in the case where $\epsilon>1$ for both 
Fermi surfaces. However, this suggests an interesting situation in the case where $\epsilon$ is greater than $1$ for only one of the Fermi surfaces i.e.  $(\epsilon_+-1)(\epsilon_- -1)<0$.
In this case, only one gap closing in the Andreev spectrum according to Eq.~\ref{eq:band_inversion} and it seems that there is no enclosed regime in the topological phase diagram in Fig.~\ref{fig:2}. 
In fact, in this case, the other topological phase boundary appears through a different mechanism. 
For $(\epsilon_+-1)(\epsilon_- -1)<0$ and $\phi_1=\pi$, let us assume without loss of generality $\epsilon_-<1$ and $\epsilon_+>1$. In this case, 
based on Eq.~\ref{eq:Dpm} the Fermi surfaces have opposite signs of pairing i.e. $\Delta_{1,-}<0$ and $\Delta_{2,+}>0$.
This is precisely the condition~\cite{teo2010topological,zhang2013time} for realizing a topologically non-trivial superconducting phase 
in the time-reversal invariant case (i.e. class DIII)~\cite{schnyder2008classification,kitaev2001unpaired,qi2010topological} in the Andreev limit 
(i.e. $\mu\gg \Delta_0$).
Such a DIII class topological phase is characterized by helical Majorana edge states appearing at the boundaries of the 2DEG
that are detectable, in principle, through quantized thermal transport at the edges.
The existence of these helical Majorana edges provides the additional TS phase boundary in the case  $(\epsilon_+-1)(\epsilon_- -1)<0$ 
along the lines $\phi_{1,2}=\pi$ in Fig.~\ref{fig:2}, which are necessary to close the boundary of the TS phase in this case.

\section{Material realization in mercury telluride}
\label{sec:4}
While the Dresselhaus model discussed in Sec.~\ref{sec:2} provides a qualitative model for generating spatially and momentum separated Fermi surface states, 
this effect is not large in conventional material in the limit of validity of the single-band model Eq.~\ref{eq:full_dresselhaus}. 
For this reason, we present results for topological superconductivity in devices where the semiconductor used in the set-up in Fig.~\ref{fig:2} is HgTe. HgTe 
has been succcessfully used in a number of Josephson junction devices~\cite{wiedenmann20164,hart2017controlled,ren2019topological} and is a semiconductor 
whose spin-orbit coupled bandstructure is very well characterized~\cite{novik2005band}.

\subsection{$8$-band model near the $\Gamma$ point}
The Bloch wave functions of HgTe at the $\Gamma$ point  are eigenstates of total angular momentum $j=1/2$ or $j=3/2$ due to spin-orbit interaction. This allows one to describe the states near the Gamma point by an $8$-band Kane model~\cite{winkler2003spin}.   Following Ref.~\cite{novik2005band}, under the $k\cdot p$ frame work, we can assume the eigenstates to take the form
\begin{equation}
    \psi_{k_{\parallel}}(\vec{r})=\sum_n F^{k_{\parallel} }_n(\vec{r})u_n(\vec{r})
\end{equation}
where the envelope function $F^{k_{\parallel}}_n(\vec{r}) = \exp{(i\vec{k}_\parallel\cdot \vec{r})}f^{k_{\parallel}}_n(z)$ can be separated into a plane wave and a $z$ dependent part and $u_n$ are the eigenstates at $\Gamma$ point where n indexes the basis states of $j=1/2$ and $j=3/2$ that are formed by the s and p orbitals.
In the presence of an external electric field $\mathcal{E}$, the wave function in the $z$ direction satisfies
\begin{equation}
    \sum_n \left[H(k_\parallel)_{mn} + V(z)\delta_{m,n}\right]f^{k_\parallel}_n(z) = Ef^{k_\parallel}_m(z)
    \label{eq:fz}
\end{equation}
where $H(k_\parallel)$ is an 8x8 matrix derived in Ref.~\cite{novik2005band} and $V(z) = e\mathcal{E} z$ is the linear potential from the electric field.
\subsection{Fermi-surface projected superconducting pairing}
As mentioned in Sec.~\ref{subsec:3B}, the proximitized pairing potential is proportional to the tunneling amplitude across the interface. We can therefore assume the superconducting term in the HgTe to take the form
\begin{equation}
    \Delta\sum_{j\in\{S,X,Y,Z\}}\int dxdy \partial_z\Psi^\dagger_{j\uparrow}(\vec{r})\bigg\rvert_{z = z_0} \partial_z\Psi^\dagger_{j\downarrow}(\vec{r})\bigg\rvert_{z = z_0} + h.c.
    \label{eq:pairing}
\end{equation}
where $\Delta=(\Delta_0/d^3)$ and $S,X,Y,Z$ represent the four atomic orbitals for $l=0$ and $l=1$ at the valence band edge and $z_0$ is the $z$ coordinate of the interface. In the eigenbasis of Eq.~\ref{eq:fz}, the field operator can be written as
\begin{equation}
    \Psi^\dagger_{j\sigma}(\vec{r}) = \sum_{m,k_\parallel} c^\dagger_{k_\parallel, m} e^{i\vec{k}_\parallel\cdot\vec{r}}\sum_n f^{(k_\parallel, m)}_n(z) C^n_{j\sigma}
    \label{eq:transformation}
\end{equation}
where $m$ is the band index and $C^n_{j,\sigma}$ is the Clebsh-Gordon coefficient of state $(j, \sigma)$ in state $n$. 
Plugging-in Eq.~\ref{eq:transformation} into Eq.~\ref{eq:pairing} gives the Cooper pairing between the eigenstates,
\begin{align}
    &\Delta\sum_{k_\parallel,m,m'} c^{\dagger}_{k_\parallel,m}c^{\dagger}_{-k_\parallel,m'}\nonumber\\ &\sum_{j\in\{S,X,Y,Z\}}\left[\sum_{n,n'} \partial_zf^{(k_\parallel, m)}_n(z_0)  \partial_zf^{(-k_\parallel, m')}_{n'}(z_0) C^n_{j\uparrow}C^{n'}_{j\downarrow}\right]
\end{align}
In the limit where the pairing potential is much smaller than the band splitting, we can neglect the off-diagonal terms in $m, m'$. Considering the contributions from the top and bottom superconductors, the pairing Hamiltonian is given by
\begin{equation}
    H_{sc} = \sum_{m, k} \Delta_m(k) c^\dagger_{k,m}c^{\dagger}_{-k,m} + h.c.
\end{equation}
where we have dropped the $\parallel$ subscript and will only work in 2D later on. The $k$-dependent pairing potential for $m$ band, $\Delta_m(k)$, exhibits p-wave symmetry,
\begin{equation}
    \Delta_m(k) = \Delta \left[\epsilon_{k,m}(0) +e^{i\phi}\epsilon_{k,m}(d)\right]
\end{equation}
where the thickness is assumed to be $d$ and the weight $\epsilon$ is given by
\begin{align}
    &\epsilon_{k, m}(z) = \nonumber\\ &i \text{Im}\left[\sum_{j\in\{S,X,Y,Z\}}\left(\sum_{n,n'} \partial_zf^{(k, m)}_n(z)  \partial_zf^{(-k,m)}_{n'}(z) C^n_{j\uparrow}C^{n'}_{j\downarrow}\right)\right]
    \label{eq:proximity_strength}
\end{align}
The imaginary part is taken as contributions from $k$ and $-k$ are combined. 
\subsection{Topological superconductivity in HgTe}
For simulation, we pick parameters $d = 10$nm and gate voltage $Ed = 0.05$V and solve for the eigenstates $f_n^{(k, m)}(z)$ in Eq.~\ref{eq:fz}. The other band structure parameters of the 8-band model are taken directly from Ref. ~\cite{novik2005band}. 

\begin{figure}
    \centering
    \includegraphics[width = 0.45\textwidth]{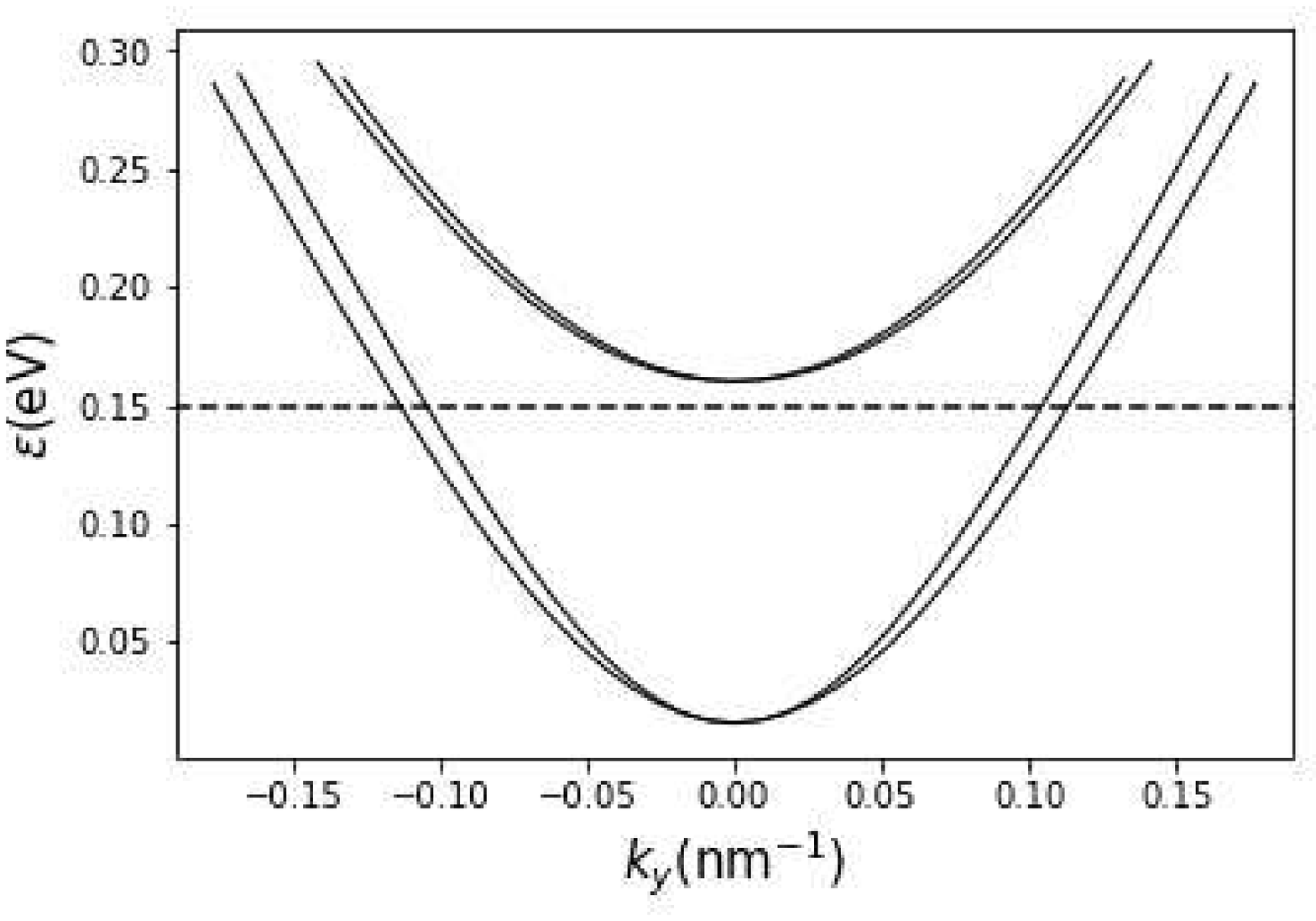}
    \caption{Dispersion relation along $k_x = 0$ of the $\Gamma_8$ bands of a slab of HgTe under external electric field. The thickness $d$ is 10nm and the electric field $E$ is set to be $Ed=0.05\text{V}$.}
    \label{fig:3}
\end{figure}

\begin{figure}
    \centering
    \includegraphics[width = 0.4\textwidth]{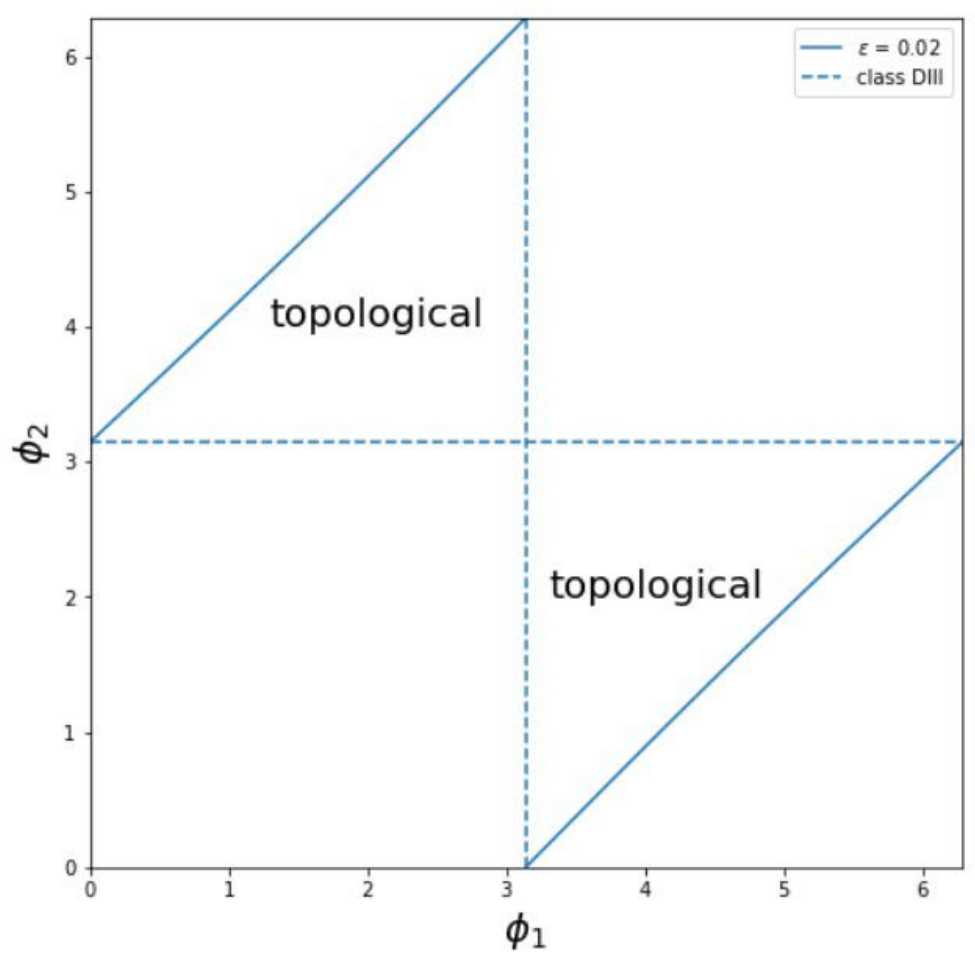}
    \caption{Topological phase diagram of HgTe with proximitized superconductivity under external electric field. The parameters are chosen to be $d = 10$nm and $Ed=0.05$V. The solid line represents the phase boundary arising from gap closing on the top surface similar to Fig.~\ref{fig:2}, and the dashed line represents that of the class DIII topological superconductor mechanism.}
    \label{fig:4}
\end{figure}

Without external field, the 8-band Hamiltonian is doubly degenerate throughout the Brillouin zone due to the combined symmetry of time-reveral and spatial-inversion symmetries $\mathcal{PT}$. Upon applying an electric field together with the infinite-well potential in the z-direction, the degeneracy is lifted. The dispersion relation for the four $\Gamma_8$ bands along $k_x = 0$ is shown in Fig.~\ref{fig:3}.
We next choose the chemical potential $\mu=0.15$ eV to have a pair of inner ($-$) and outer ($+$) Fermi surfaces. 
By plugging the wave functions into Eq.~\ref{eq:proximity_strength}, we get the strength $\epsilon_{k,\pm}$ of the proximity effect from the top and bottom Fermi surfaces.
The resulting relative proximitized pairing amplitude $\epsilon_{\pm}$ is
\begin{align}
    \epsilon_- \approx 12.70 \nonumber\\
    \epsilon_+ \approx 0.01773
    \label{eq:relative_proximity_effect}
\end{align}
The $\epsilon_{\pm}$ satisfies $(\epsilon_+ - 1)(\epsilon_- - 1)<0$. As discussed in Sec.~\ref{subsec:3C}, one side of the the topological regime is enclosed by the band inversion points in Andreev spectrum according to Eq.~\ref{eq:band_inversion}  
shown in Fig.~\ref{fig:2} and the other is from the class DIII mechanism which takes place at precisely $\phi_{1,2} = \pi$. The resulting phase diagram is shown in Fig.~\ref{fig:4}.
Incidentally, this phase diagram appears essentially identical to what one would expect from the 3D TI device shown in the upper panel of Fig.~\ref{fig:1}.
This can be understood from Eq.~\ref{eq:relative_proximity_effect}, which shows that despite not being a 3D TI, HgTe in the presence of an electric field creates states that are essentially localized on the top and bottom surfaces much like a 3D TI thin film.

\subsection{Andreev spectrum at finite $k_x$}
\label{subsec:4D}
\begin{figure}[t]
    \centering
    \includegraphics[width = 0.4\textwidth]{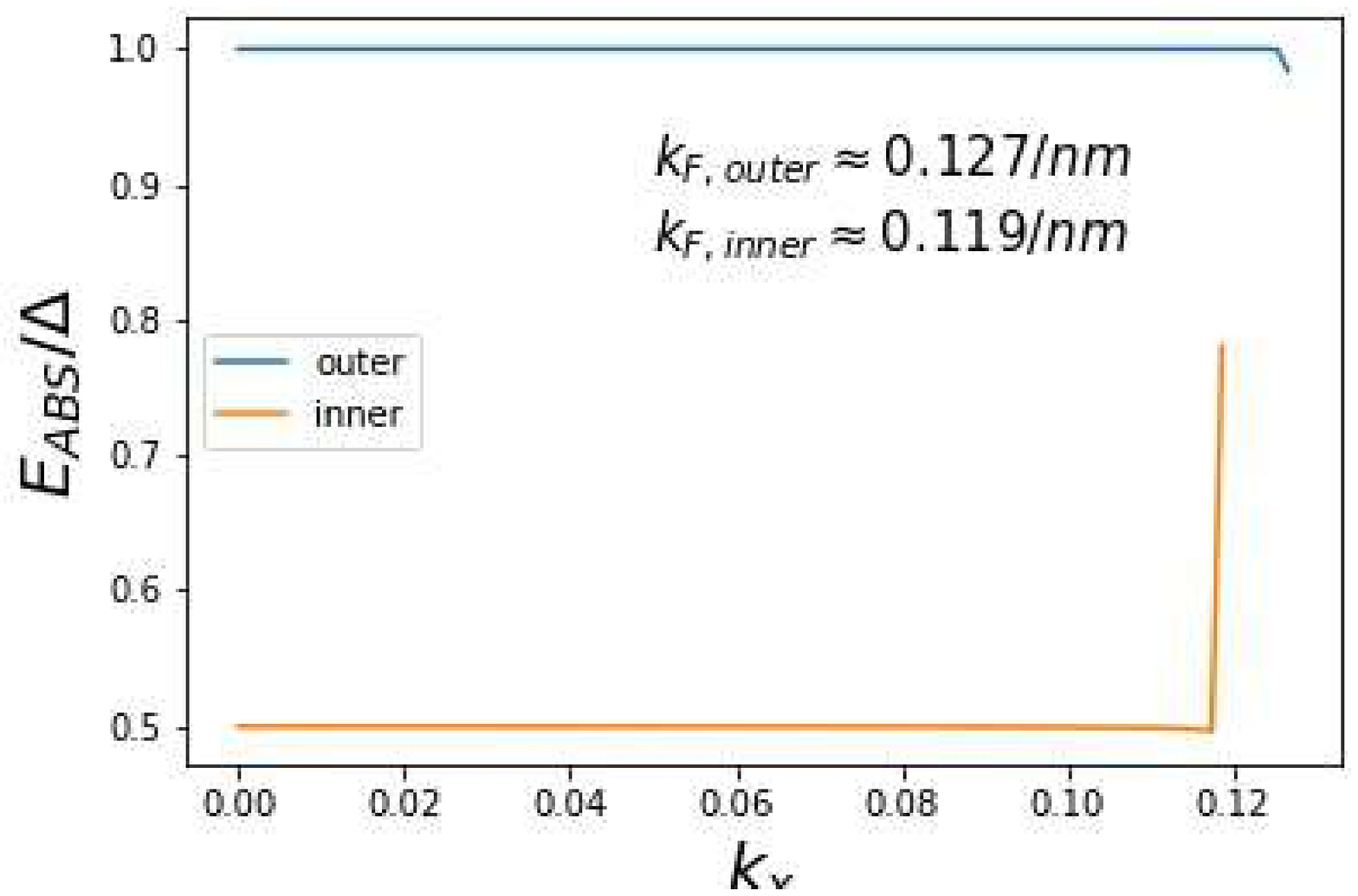}
    \caption{Energy gap of the Andreev spectrum as a function of $k_x$. The parameters are $\Delta_0 = 1\text{K}, m^*_{+(-)} = 0.0534(0.0469)m_e$, $\phi_{+(-)} = 0.01(4\pi/3)$ and Josephson junction width $d = 10\text{nm}$}
    \label{fig:5}
\end{figure}

For the MBSs to be robust, the planar Josephson junction must be gapped throughout the Fermi surface. 
To verify this is true in our system, we simulate the Andreev spectrum at finite $k_x$. 
As an approximation, we will consider the dispersion of the two low energy bands in Fig~\ref{fig:3} to be quadratic, or $\epsilon(\vec{k}) \approx (k_x^2 +k_y^2)/2m^*$. 
The Andreev spectrum can be obtained by matching the continuous boundary condition, as described in appendix~\ref{app.A}, with the effective chemical potential shifted to $\mu' = \mu - k_x^2/2m^*$. 
In Fig. ~\ref{fig:5}, we show an example of the ABS energy gap along the Fermi surface. With chemical potential fixed to $ \mu = 0.15\text{eV}$, we choose the pairing potential $\Delta_0 = 1\text{K}$, Josephson junction width $w = 100\text{nm}.$ 
The phase differences for the two Fermi surfaces are set to be $\phi_{+(-)} = 0.01(4\pi/3)$ and the effective masses $m^*_{+(-)} = 0.0534 (0.0469)m_e$ are obtained through fitting the curvatures at band minima in the dispersion shown in Fig.~\ref{fig:3}.
Note that although there are visible discrepancies between the dispersion in Fig.~\ref{fig:3} and the quadratic fit at $k_y$ closer to $k_F$, the ABS gap in Fig.~\ref{fig:5} at large $k_x$ is expected to be accurate since we are effectively looking at the band minimum. 
On the other hand, at small $k_x$, the system is well approximated by the Andreev limit, where the gap does not depend on details of the dispersion of the energy band.
This qualitatively justifies our approximation of the quadratic dispersion for the ABS gap calculation.

\section{Summary and discussion}
In this work, we 
show that a planar Josephson junction with superconductors on both surfaces  provide a way to create MZMs without an external magnetic field in a 
single JJ. Such devices show intriguing similarities to MZMs in a 3DTI~\cite{fu2008superconducting} and can therefore take advantage 
of strong spin-orbit coupled materials such as HgTe that are in the vicinity of a 3D TI phase.
We showed that a 2DEG composed of a semiconductor with bulk inversion asymmetry i.e. with a cubic Dresselhaus term can lead to Fermi surfaces that are split both in momentum space as well as between the top and bottom surfaces in the presence of a vertical electric field. The spatial separation is a key ingredient that allows one to introduce different superconducting phases in the top and bottom surfaces. 
The essential ingredient for differentiating the effective pairing phase between the two Fermi surfaces is a coupling between $k_z$ and $k_\parallel$, which in this model is generated by the cubic Dresselhaus interaction. 
We showed that a stronger effect occurs in a realistic model for a HgTe 2DEG, though the origin of the spatial separation is more complex owing to the multiband Hamiltonian. In fact, it is possible that the electric field effectively converts HgTe into a 3D TI thin film in a way that is analogous to the strain effect~\cite{brune2011quantum}. This highlights the advantage of the proposed device in Fig.~\ref{fig:1}, which can show a topological phase for a range of spin-orbit coupling models from conventional Dresselhaus to those in 3D TIs.

One caveat that should be emphasized is that  the present work uses a rather simple model of a uniform electric field across the 2DEG, which may be difficult to apply in the presence of a superconductor. We expect that choosing the width of the JJ in Fig.~\ref{fig:1} to be comparable to the coherence length should be sufficient to ensure that the electric field in the effective superconducting phase is controlled by the vertical profile of the wave-function in the JJ. One expects the effect of the electric to be screened underneath the SCs. However, in the limit of strong coupling between the 2DEG and the SC, the wave-functons in the 2DEGs can be assumed to have rather short coherence lengths~\cite{sau2010robustness}. A precise estimate of the effect of screening will depend on the details of the potential at the semiconductor/superconductor interface as well as requiring a self-consistent solution of the Schrodinger-Poisson problem. This could be of interest in a future work modeling a device made from a specific set of materials.  

This work was supported by the Joint Quantum Institute, NSF DMR 1555135 (CAREER) and JQI-NSF-PFC (NSF Grant No. PHY-1430094). JS acknowledges valuable 
conversations with Nitin Samarth that motivated the study of semiconductor/superconductor sandwich structures as well as with Yuval Oreg regarding the 
literature on phase biased Josephson junctions.

\bibliography{apssamp}

\newpage

\appendix
\section{Details of Dresselhaus solution}
\label{app.0}
The Schrodinger equation for each sector, after subtracting $p^2/2m\pm D p^3\sqrt{2}$ is 
\begin{align}
    &E_\pm \phi =  [p_z^2(1/2m\mp D\sqrt{2}p)+V(z)]\phi(z).
\end{align}
Scaling $z\rightarrow \lambda z$, $p_z\rightarrow \lambda^{-1}z$ by a factor $\lambda$ so that $Q=\lambda^{-2}(1/2m\mp D\sqrt{2}p)=\mathcal{E}\lambda$ changes the Schrodinger equation to 
\begin{align}
    &E_\pm \phi = Q [p_z^2+z]\phi(\lambda z),
\end{align}
where $\lambda^3=\mathcal{E}^{-1}(1/2m\mp D\sqrt{2}p)$ and $Q=\mathcal{E}\lambda=\mathcal{E}^{2/3}(1/2m\mp D\sqrt{2}p)^{1/3}$. The energy eigenvalue 
\begin{align}
    &E_{\pm}=\nu_0 Q\nonumber\\
    &\phi(z)=\nu_1\lambda^{-1/2}Ai(\lambda^{-1}z-\nu_2),
\end{align}
where $\nu_{j=0,1,2}$ are numerical factors associated with Airy's differential equation. 
Restoring the energy shift, the energy eigenvalues are 
\begin{align}
&E_\pm=p^2/2m\pm D p^3\sqrt{2}+ \alpha \mathcal{E}^{2/3}(1/2m\mp D\sqrt{2}p)^{1/3}\nonumber\\
&\simeq p^2/2m\pm D p^3\sqrt{2}+ \alpha \mathcal{E}^{2/3}(1/2m\mp  D\sqrt{2}p/3).
\end{align}

\section{Andreev bound state energy close to the band minimum}
\label{app.A}
We consider a spinless 1d s-n-s Josephson junction described by the Hamiltonian
\begin{equation}
    H(x) = H_0 + \left[\Theta(x-d/2)e^{i\phi/2} + \Theta(-d/2-x)e^{-i\phi/2}\right]H_{sc},
\end{equation}
where $d$ is the width of the junction and $\phi$ is the superconducting phase difference across the junction. Close to the band minima, we assume quadratic dispersion for $H_0$,
\begin{align}
    H_0 &= \int_x \Psi^\dagger(x)(-\partial^2_x/2m-\mu)\Psi(x)\nonumber\\
    H_{sc} &= \int_x\int_y \Delta(x-y)\Psi^\dagger(x)\Psi^\dagger(y) + h.c.
\end{align}
On the two superconductor sides, the wave function in the spinor representation $\gamma_k^\dagger = u_k c_k^\dagger + v_k c_{-k}$ satisfies the Bogoliubov-de Gennes equation given by
\begin{equation}
    \begin{pmatrix}
    \frac{k^2}{2m}-\mu & \Delta(k) \\
    \Delta^*(k) & -\frac{k^2}{2m}+\mu
    \end{pmatrix}
    \begin{pmatrix}
    u_k\\
    v_k
    \end{pmatrix}
    =
    E
    \begin{pmatrix}
    u_k\\
    v_k
    \end{pmatrix}
\end{equation}
where $\Delta(k) = \Delta_0 e^{-i\theta_k}$ from Eq.~\ref{eq:p-wave}. This gives the relation between the momentum $k$ and energy $E$
\begin{equation}
    k = \pm \sqrt{2m\left(\mu\pm\sqrt{E^2-\Delta_0^2}\right)}.
\end{equation}
The Andreev bound states (ABSs) are in-gap states ($\vert E\vert<\Delta_0$) whose wave function decays exponentially within the superconductors.
That is, for $x > d/2$, $k = \kappa$ or $-\kappa^*$ and for $x < -d/2$, $k = -\kappa$ or $\kappa^*$, where 
\begin{equation}
    \kappa = \sqrt{2m\left(\mu+\sqrt{E^2-\Delta_0^2}\right)}
\end{equation}
and in the middle ($-d/2<x<d/2$), $k = \pm q_e$ or $\pm q_h$, where
\begin{align}
    q_e &= \sqrt{2m(\mu+E)}\nonumber\\
    q_h &= \sqrt{2m(\mu-E)}.
\end{align}
Consequently, the wave function in the three regions satisfies the general form
\begin{align}
\label{eq:continuity}
    \vert \psi_L\rangle &= A_L
    \begin{pmatrix}
    e^{-i\phi/2}\Delta(-\kappa) \\
    E-\sqrt{E^2-\Delta_0^2}
    \end{pmatrix}
    e^{-i\kappa x}
    \nonumber\\ &+ B_L
    \begin{pmatrix}
    e^{-i\phi/2}\Delta(\kappa^*) \\
    E+\sqrt{E^2-\Delta_0^2}
    \end{pmatrix}
    e^{i\kappa^* x} \nonumber\\ 
    \vert \psi_M\rangle &= 
    \begin{pmatrix}
    A_Me^{iq_ex}+B_Me^{-iq_ex} \\
    C_Me^{iq_hx}+D_Me^{-iq_hx}
    \end{pmatrix}
    \nonumber \\ 
    \vert \psi_R\rangle &= A_R
    \begin{pmatrix}
    e^{i\phi/2}\Delta(\kappa) \\
    E-\sqrt{E^2-\Delta_0^2}
    \end{pmatrix}
    e^{i\kappa x}
    \nonumber\\& + B_R
    \begin{pmatrix}
    e^{i\phi/2}\Delta(-\kappa^*) \\
    E+\sqrt{E^2-\Delta_0^2}
    \end{pmatrix}
    e^{-i\kappa^* x}
\end{align}
By imposing the continuity condition at the boundaries $x = \pm d/2$, one can solve for the Andreev spectrum. 

In the Andreev limit $\xi_0 \gg \lambda_F$ or equivalently, $\mu \gg \Delta_0$, the set boundary condition for the states in Eq. \ref{eq:continuity} is particularly simple, and the ABS energy is given by the usual form~\cite{sauls2018andreev}
\begin{equation}
    \frac{E}{\Delta_0} = \pm \cos\left(\bar{w}\frac{E}{\Delta_0}+\frac{\phi}{2}\right),
    \label{eq:ABS0}
\end{equation}
where $\bar{w} = w/(\hat{p}\cdot\hat{n}\xi_0)$ is the effective width measured by the coherence length $\xi_0 = v_F/\Delta_0$.

\section{Effective $p-$wave pairing}
\label{app.A.5}
Due to the z confinement, we can assume $\langle p^2_z\rangle \gg \langle p^2_x\rangle, \langle p^2_y\rangle$ in the low energy limit. 
In this case, the leading order effect is the linear Dresselhaus effect in the $x-y$ plane.
\begin{equation}
    H^{(0)} = \frac{1}{2m}\left(p_x^2+p_y^2\right)+\frac{\beta}{\hbar} \left(-p_x\sigma_x + p_y\sigma_y\right)
\end{equation}
The eigen-energy is given by
\begin{equation}
\varepsilon_\pm(\vec{p}) = \frac{p^2}{2m} \pm \frac{\beta}{\hbar} p
\end{equation}
\begin{equation}
\begin{pmatrix}
c^\dagger_{k,+}\\
c^\dagger_{k,-}
\end{pmatrix}
=
\frac{1}{\sqrt{2}}\begin{pmatrix}
e^{i\theta_k} & 1\\
e^{i\theta_k} & -1
\end{pmatrix}
\begin{pmatrix}
c^\dagger_{k,\uparrow}\\
c^\dagger_{k,\downarrow}
\end{pmatrix}
\label{eq:soc_fermion}
\end{equation}
where $\theta_k = \tan^{-1}\left(k_y/k_x\right)$ is the polar angle of momentum $k$ in the $k_x$-$k_y$ plane.

Next, we consider a proximitized s-wave superconductivity,
\begin{equation}
    H_{sc} = \Delta\sum_k c^\dagger_{k,\uparrow}c^\dagger_{-k,\downarrow} + h.c.
    \label{eq:s-wave_pairing}
\end{equation}
Plugging in Eq.~\ref{eq:soc_fermion} into Eq.~\ref{eq:s-wave_pairing} leads to a pairing potential that takes the $p+ip$ form.
\begin{align}
    H_{sc} =\frac{\Delta}{2}\sum_{k} e^{-i\theta_k}\left(c^\dagger_{k,+}c^\dagger_{-k,+}-c^\dagger_{k,-}c^\dagger_{-k,-}\right)+h.c.
    \label{eq:p-wave}
\end{align}
As a result, in the presence of a proximitized s-wave pairing, the split bands ($\pm$) from Dresselhaus spin-orbit coupling exhibit spinless p-wave superconductivity separately.

\section{Andreev spectrum between superconductors of different pairing amplitude}
\label{app.B}
As discussed in Ref.~\cite{beenakker1991universal, titov2006josephson}, the Andreev bound state spectrum in an SNS junction can be obtained through the scattering theory of the BdG Hamiltonian with the step function pairing potential. 
In the Andreev limit $\mu \gg \vert\Delta\vert$ and considering the bound state regime $E < \vert \Delta \vert$, pure Andreev reflection happens at an ideal NS interface. The reflection coefficients can be obtained by considering the continuous boundary conditions at the $0^{th}$ order in $E/\mu$ and $\vert \Delta\vert/\mu$, which are given by
\begin{align}
\label{eq:andreev_reflection}
    r_{ee} &= 0 \nonumber\\
    r_{hh} &= 0 \nonumber\\
    r_{he} &= e^{-i\left(\phi+\beta\right) }\nonumber\\
    r_{eh} &= e^{i\left(\phi-\beta\right)}
\end{align}
where $\beta = \text{arccos}(E/\vert\Delta\vert)$ and $\phi$ is the order parameter phase of the superconductor.
On the other hand, across the normal metal, the transmission amplitude is simply the phase acquired by the plane wave $t_{e/h}(k,\Delta y) = \text{exp}\left(\pm ik\Delta y\right)$, and the momentum of the electron and hole branch can be approximated by the linear relation $k_{e/h}(E)\approx k_F \pm E/v_F$.
The quantization condition is that the transmission amplitude of a round trip equals unity. That is,
\begin{align}
t_{\circlearrowleft}= r^R_{he}t_h\left(-k(E), -w\right)r^L_{eh}t_e\left(k(E),w\right) = 1.
\end{align}
Assuming the order parameters on the left/right superconductors to be $\Delta_{L/R} e^{i\phi_{L/R}}$, the ABS energy satisfies
\begin{equation}
    \label{eq:ABS1}
    2w\frac{E}{v_F} - \text{arccos}\left(E/\Delta_L\right)-\text{arccos}\left(E/\Delta_R\right) = 2n\pi + \Delta\phi
\end{equation}
where $\Delta\phi \equiv \phi_R - \phi_L$.
Note that Eq.~\ref{eq:ABS1} can be reduced to Eq.~\ref{eq:ABS0} when taking the superconducting strength to be equal on both sides $\Delta_L = \Delta_R = \Delta_0$,

An important property of Eq.~\ref{eq:ABS1} is that at $\Delta\phi=\pi$ there is always a zero energy ABS. This band inversion point $\phi=\pi$ is independent of the relative superconducting strength of the two superconductors. As a result, the topological phase boundary of the planar Josephson junction can be determined solely by the effective phase difference across the Josephson junction for the two Fermi surfaces at the symmetry point $k_x = 0$.

\end{document}